\documentclass[12pt]{article}

\begin{document}

\centerline{\Large Time evolution of Hamiltonian constraint system:}
\centerline{\Large an idea applicable to quantum gravity} 
\vspace{0.5cm} 
\centerline{\large\it Junichi Iwasaki}
\vspace{0.3cm}
\centerline{iwasaki@physicist.net}
\vspace{0.3cm}
\centerline{Oct 26, 2012}


\begin{abstract}
The Hamiltonian constraint system is the canonical formulation of a physical
system with a Hamiltonian constrained to vanish.
In terms of the canonical variables, we define what we call reference
observable, with respect to which other observables evolve.
We study if it plays the role of time.
As simple examples, we study the theories of non-relativistic and relativistic 
particles.
We outline an application of the idea to general relativity.
\end{abstract}


\section{Introduction}\label{sec:intro}

The Hamiltonian constraint system is the canonical formulation of a physical
system with a Hamiltonian constrained to vanish.
This Hamiltonian generates the evolution of observables 
with respect to a parameter given in the theory.
This parameter is arbitrarily reparametrizable and hence it is
not physical time but a gauge parameter.
Physical time is given as an observable in some theories or
not given explicitly in the other theories.
In terms of canonical variables, we define what we call reference
observable, with respect to which other observables evolve.
We study if the reference observable plays the role of time.

General relativity (GR) is a Hamiltonian constraint system.
The time coordinate given in GR is not physical time. 
Physical time is not explicitly given in the theory.
Therefore, a reference observable, if defined, is perhapes
utilized as physical time.
The idea of reference observable already exists although not 
clearly realized. 
It is in the theory of relativistic particle (RP),
a Lorentz covariant theory, which is 
another Hamiltonian constraint system much simpler than GR. 
RP is a trajectory theory while GR is a field theory.
Its physical time is one of the observables although it is not
Lorentz invariant. 
Lorentz invariant time is not given explicitly in the theory.
{}Furthermore, The theory of non-relativistic particle (NP),
a Galilei covariant theory, is
known to be written as a Hamiltonian constraint system,
another trajectory theory simpler than RP.
Its physical time is one of the observables and Galilei invariant.

NP is a toy we use to study the idea in Sec.\ref{sec:non}.
We see the reference observable we define for NP is 
the physical time, which is Galilei invariant. 
Upon quantization, we see a Heisenberg picture with respect to
the referene observable is available.

We extend the idea to RP in Sec.\ref{sec:rel}.
We see the reference observable we define for RP is 
the proper time, which is Lorentz invariant.
We also see that a Heisenberg picture with respect to
the reference observable can be constructed in quantization.

The definition of the reference observsble is not 
particular to the theories studied here but general enough 
for the Hamiltonian constraint system.
We outline an application of the idea to GR in Sec.\ref{sec:gen}.
We conclude in Sec.\ref{sec:con}.

The present work is motivated by the use of the Heisenberg picture
rather than the Schr\"odinger picture \cite{CR1992}
and relational spacetime rather than absolute spacetime \cite{CR2004}
in quantization of non-perturbative canonical gravity \cite{AA1991}.


\section{Non-relativistic particle (NP)}\label{sec:non}
\subsection{Gauge independent formulation}\label{sub:indep}

NP or more precisely the non-relativistic trajectory of a free particle 
is often described as a Hamiltonian system without constraint.
The action is 
\begin{eqnarray}
&&
S_0:=\int\left(p{dx\over dt}-H_0\right)dt,
\label{eq:S_0}
\end{eqnarray}
where $x$ and $p$ are the position and momentum of 
the particle respectively,
$t$ is physical time and 
\begin{eqnarray}
&&
H_0:={p^2\over 2m}
\label{eq:H_0}
\end{eqnarray}
is the Hamiltonian.
$m$ is the mass of the particle.
Here the space is assumed to be one dimensional for simplicity.

$x$ and $p$ are the canonical variables of the Hamiltonian system
and satisfy the Poisson bracket relations 
$\{x,p\}=1$ and $\{x,x\}=\{p,p\}=0$.
They are functions of $t$ and their evolution in $t$ is generated by 
the Hamiltonian.

The equations of motion are
\begin{eqnarray}
&&
{dx\over dt}=\{x,H_0\}={p\over m},
\label{eq:dxdt}
\\
&& 
{dp\over dt}=\{p,H_0\}=0.
\label{eq:dpdt}
\end{eqnarray}
This system has no constraint nor gauge parameter.
The equations of motion above present physics.
We refer this system to gauge independent formulation and
the equations of motion above to gauge independent equations of motion.

Quantization is a process of 
replacing $p$ by $\hat p:=-i{\partial\over\partial x}$
and the Poisson braket by $(-i)$ times the commutator
in a representation space in which $x$ is diagonal.
The state function is a function of $x$ and $t$.
By applying the time evolution operator to a Fourier mode,
the physical state with momentum $p$ is 
\begin{eqnarray}
&
\psi_0(x,t)
&
=e^{-i{\hat p^2\over 2m}t}e^{ipx}
=e^{i\left(px-{p^2\over 2m}t\right)}
\label{eq:psi_0_S}
\\
&&
=e^{i\int_0^p\left(x-{k\over m}t\right) dk}.
\label{eq:psi_0_H}
\end{eqnarray}
Here, $x$ is the position of the particle at time $t$ and,
when moving with momentum $p$, 
$x-{p\over m}t$ is the position at $t=0$.
Therefore, (\ref{eq:psi_0_H}) is in the Heisenberg picture 
while (\ref{eq:psi_0_S}) is in the Schr\"odinger picture.
Since the classical solutions are labelled by the initial position
and momentum, the physical state in the Heisenberg picture is seen
as a function of the classical solutions.

The gauge independent formulation for NP summarized in this
subsection is well understood and 
a good toy for us to study the idea introduced in this paper.

\subsection{Gauge covariant formulation}\label{sub:covar}

NP is restructured as a Hamiltonian constraint 
system.
The action is
\begin{eqnarray}
&&
S_N:=\int\left(p{dx\over d\tau}+e{dt\over d\tau}-H_N\right)d\tau,
\label{eq:S_N}
\end{eqnarray}
where $\tau$ is an arbitrary parameter and
\begin{eqnarray}
&&
H_N:=\lambda\left(e+{p^2\over 2m}\right)
\label{eq:H_N}
\end{eqnarray}
is the Hamiltonian. 
$\lambda$ is a multiplier to force $e=-{p^2\over 2m}$.
Physical time $t$ is treated as another canonical variable with its canonical
conjugate $e$ satisfying $\{t,e\}=1$ and $\{t,t\}=\{e,e\}=0$.
The canonical variables are now functions of $\tau$ and their evolution 
in $\tau$ is generated by the Hamiltonian, which is constrained to vanish.

Note that by substituting $e=-{p^2\over 2m}$, $S_0$ is restored from $S_N$
provided $t$ is a monotonic function of $\tau$.
However, without doing so, we restore physical content of the gauge
independent formulation in Sec.\ref{sub:galil}.

The equations of motion together with the constraint are
\begin{eqnarray}
&&
\dot{t}=\{t,H_N\}=\lambda,
\label{eq:tdot}
\\
&&
\dot{x}=\{x,H_N\}=\lambda{p\over m},
\label{eq:xdot}
\\
&&
\dot{e}=\{e,H_N\}=0,
\label{eq:edot}
\\
&&
\dot{p}=\{p,H_N\}=0,
\label{eq:pdot}
\\
&&
e+{p^2\over 2m}=0.
\label{eq:e+}
\end{eqnarray}
Here the dot implies the derivative with respect to $\tau$.
In the phase space spanned by the canonical variables,
we refer the subspace satisfying the constraint (\ref{eq:e+})
to the constraint surface.
Because $\tau$ is arbitrarily reparametrizable, the equations of motion
above are gauge dependent.
Rather, they are gauge covariant since the change of $\lambda$ together
with a reparametrization of $\tau$ results in the same equations in form.
We refer them to gauge covariant equations of motion.
Accordingly, we refer the Hamiltonian constraint system to
gauge covariant formulation.
As we see in the following sections, RP and GR are in this kind of systems.

\subsection{Galilei invariant reference observable}\label{sub:galil}

In this subsection, we define what we call reference observable.
Then, using it, we derive the gauge independent equations of motion
from the gauge covariant ones.

We denote the reference observable by $s$. 
The definition is such that it is an observable satisfying
$\dot s=\lambda$ on the constraint surface.
$t$ satisfies this condition and hence can be a reference observable.
However, we look for a more general form of $s$ so that the idea can be 
applied to RP and GR, that is, $s$ containing terms of higher order 
in canonical variable.
It can be 
\begin{eqnarray}
&&
s:=t-c\left({te\over m}+{xp\over 2m}\right),
\label{eq:s}
\end{eqnarray}
where $c$ is a constant, 
so that 
\begin{eqnarray}
&&
\dot s=\{s,H_N\}=\lambda-{c\over m}H_N.
\label{eq:sdot}
\end{eqnarray}
The second term in (\ref{eq:sdot}) vanishes on the constraint surface. 
This means that the second term in (\ref{eq:s}) is constant in $\tau$ 
on the constraint surface.

In the gauge covariant formulation, the canonical variables  
are functions of $\tau$.
Here, we restrict ourselves to a sector in which the canonical variables are
functions of $s$, through which they implicitly depend on $\tau$,
that is, $t(\tau):=t(s(\tau))$, $x(\tau):=x(s(\tau))$, $e(\tau):=e(s(\tau))$ 
and $p(\tau):=p(s(\tau))$.
This sector contains all the solutions for the gauge independent 
equations of motion.
If $\lambda>0$, then the sector is identical to the space of 
the canonical variables we started with.

We use 
${d\over d\tau}={ds\over d\tau}{d\over ds}=\lambda {d\over ds}$
to rewrite the gauge covariant equations of motion as follows.
\begin{eqnarray}
&&
{dt\over ds}=\{t,e+H_0\}=1, 
\label{eq:dtds}
\\
&&
{dx\over ds}=\{x,e+H_0\}=\{x,H_0\}={p\over m}, 
\label{eq:dxds}
\\
&&
{de\over ds}=\{e,e+H_0\}=\{e,H_0\}=0,
\label{eq:deds}
\\
&&
{dp\over ds}=\{p,e+H_0\}=\{p,H_0\}=0.
\label{eq:dpds}
\end{eqnarray}
These equations and their solutions are independent of $\tau$.
Since equation (\ref{eq:dtds}) implies $s=t$ up to a constant term
and equation (\ref{eq:dpds}) implies (\ref{eq:deds}) 
on the constraint surface,
the gauge independent equations of motion 
(\ref{eq:dxdt}) and (\ref{eq:dpdt}) are restored.

We see the reference observable is in fact the physical time
of the theory and Galilei invariant.
It must be pointed out that the definition of $s$ here is not 
particular to the present theory but general enough for the Hamiltonian 
constraint system.

\subsection{Heisenberg picture in reference observable}\label{sub:heise}

In the gauge covariant formulation, quantization is a process of 
replacing $p$ and $e$ by $\hat p:=-i{\partial\over\partial x}$
and $\hat e:=-i{\partial\over\partial t}$ respectively and
the Poisson braket by $(-i)$ times the commutator
in a representation space in which $x$ and $t$ are diagonal.
Accordingly, the equations of motion and the constraint are replaced by
the respective representation as operators on the state function.

The state function is a function of $x$, $t$ and $s$.
However, the physical state should not depend on $s$ because it must
be equal to $\psi_0$, that is gauge independent.
The physical state with momentum $p$ is
\begin{eqnarray}
&
\psi_N(x,t)
&
=e^{-i\left(\hat e+{\hat p^2\over 2m}\right)s}
e^{i\left(px+et\right)}
=e^{i\left(px+et-\left(e+{p^2\over 2m}\right)s\right)}
\label{eq:psi_N_S}
\\
&&
=e^{i\int_0^p\left(x-{k\over m}s\right) dk}
e^{i\int_0^e\left(t-s\right) dh}
\label{eq:psi_N_H}
\end{eqnarray}
with $e+{p^2\over 2m}=0$.
Here, the operator applied makes an evolution in $s$.
$x$ and $t$ are the position and time of the particle respectively at
the reference observable $s$ and,
when moving with momentum $p$, 
$x-{p\over m}s$ and $t-s$ are the position 
and time respectively at $s=0$.
Therefore, with respect to the reference observable,
(\ref{eq:psi_N_H}) is in the Heisenberg picture 
while (\ref{eq:psi_N_S}) is in the Schr\"odinger picture
before imposing the constraint.
However, after imposing the constraint, (\ref{eq:psi_N_S}) cannot be
in the Schr\"odinger picture since the $s$ dependence in (\ref{eq:psi_N_S})
disappears as expected.
Nevertheless, the Heisenberg picture of (\ref{eq:psi_N_H}) holds because
the presence of $s$ in (\ref{eq:psi_N_H}) is not due to the state function 
itself but due to the gauge dependence of $x$ and $t$.
Hence, the reference observable $s$ behaves as if it is time
in the Heisenberg picture.


\section{Relativistic particle (RP)}\label{sec:rel}
\subsection{Gauge covariant formulation}\label{sub:r-gauge}

RP or more precisely the relativistic trajectory of a free particle 
is often described as a Hamiltonian constraint system,
which is a gauge covariant formulation in our classification.
The action is
\begin{eqnarray}
&&
S_R:=\int\left(p_\mu{dx^\mu\over d\tau}-H_R\right)d\tau,
\label{eq:S_R}
\end{eqnarray}
where $x^\mu$ and $p_\mu$ are the 4-position and 4-momentum of 
the particle respectively,
$\tau$ is an arbitrary parameter and
\begin{eqnarray}
&&
H_R:={\lambda\over 2m}\left(p_\mu p^\mu+m^2\right)
\label{eq:H_R}
\end{eqnarray}
is the Hamiltonian. 
$\lambda$ is a multiplier to force $p_\mu p^\mu+m^2=0$
and $m$ is the mass of the particle.
Here the Greek letter indices run $0$ through $3$ and are raised 
or lowered by Minkowski metric with the signature $(-1,+1,+1,+1)$
and the sum over repeated indices is understood
unless otherwise explicitly stated.

$x^\mu$ and $p_\mu$ are the canonical variables of the theory and
satisfy the Poisson bracket relations
$\{x^\mu ,p_\nu\}=\delta^\mu_{\ \nu}$ and 
$\{x^\mu ,x^\nu\}=\{p_\mu ,p_\nu\}=0$.
They are observables of the theory.
They are functions of $\tau$ and their evolution in $\tau$
is generated by the Hamiltonian,
which is constrained to vanish.
The gauge covariant equations of motion together with the constraint are 
\begin{eqnarray}
&&
\dot x^\mu=\{x^\mu ,H_R\}={\lambda\over m}p^\mu,
\label{eq:xdotmu}
\\
&&
\dot p_\mu=\{p_\mu ,H_R\}=0,
\label{eq:pdotmu}
\\
&&
p_\mu p^\mu+m^2=0,
\label{eq:pmupmu+}
\end{eqnarray}
where the dot implies the derivative with respect to $\tau$.
We refer the subspace in the phase space satisfying the constraint
(\ref{eq:pmupmu+}) to the constraint surface.

\subsection{Lorentz invariant reference observable}\label{sub:r-loren}

In this subsection, we define a reference observable.
Then, we derive the equations of motion with respect to it, that is,
the gauge independnt equations of motion.

We denote the reference observable by $s$. 
The definition is such that it is an observable satisfying
 $\dot s=\lambda$ on the constraint surface.
It is satisfied by 
\begin{eqnarray}
&&
s:=-{x^\mu p_\mu \over m},
\label{eq:ss}
\end{eqnarray}
up to a constant term.
It is straightforward to show that 
\begin{eqnarray}
&&
\dot s=\{s,H_R\}=\lambda-{2\over m}H_R.
\label{eq:ssdot}
\end{eqnarray}

In the gauge covariant formulation, the canonical variables  
are functions of $\tau$.
Here, we restrict ourselves to a sector in which the canonical variables are
functions of $s$, through which they implicitly depend on $\tau$,
that is, $x^\mu(\tau):=x^\mu(s(\tau))$, and $p_\mu(\tau):=p_\mu(s(\tau))$.
This sector contains all the solutions for the gauge independent 
equations of motion.
If $\lambda>0$, then the sector is identical to the space of 
the canonical variables we started with.

We use ${d\over d\tau}={ds\over d\tau} {d\over ds}=\lambda {d\over ds}$,
to rewrite the gauge covariant equations of motion as follows. 
\begin{eqnarray}
&&
{dx^\mu\over ds}=\{x^\mu,{{p_\nu p^\nu+m^2}\over 2m}\}
=\{x^\mu,{p_\nu p^\nu\over 2m}\}={p^\mu\over m},
\label{eq:dxmuds}
\\
&&
{dp_\mu\over ds}=\{p_\mu,{{p_\nu p^\nu+m^2}\over 2m}\}
=\{p_\mu,{p_\nu p^\nu\over 2m}\}=0.
\label{eq:dpmuds}
\end{eqnarray}
These equations are independent of $\tau$.

The reference observable $s$ for the present theory is known to be 
the proper time since 
$\sqrt{-{dx_\mu\over ds}{dx^\mu\over ds}}=
\sqrt{-{p_\mu\over m}{p^\mu\over m}}=1$,
although this definition of the proper time is not commonly discussed.
It must be pointed out that the definition of $s$ here is not particular
to the present theory but general enough for the Hamiltonian constraint system.

\subsection{Lorentz invariant Heisenberg picture}\label{sub:r-heise}

Quantization is a process of 
replacing $p_\mu$ by $\hat p_\mu:=-i{\partial\over\partial x^\mu}$
and the Poisson braket by $(-i)$ times the commutator
in a representation space in which $x^\mu$ is diagonal.

The state function is a function of $x^\mu$ and $s$.
The physical state with momentum $p_\mu$ is 
\begin{eqnarray}
&
\psi_R(x)
&
=e^{-i{\hat p_\nu \hat p^\nu+m^2\over 2m}s}e^{i p_\mu x^\mu}
=e^{i\left(p_\mu x^\mu-{p_\nu p^\nu+m^2\over 2m}s\right)}
\label{eq:psi_R_S}
\\
&&
=e^{i\int_0^p\left(x^\mu-{k^\mu\over m}s\right) dk_\mu}
e^{-i{m\over 2}s}
\label{eq:psi_R_H}
\end{eqnarray}
with $p_\mu p^\mu+m^2=0$.
Here, the operator applied makes an evolution in $s$.
$x^\mu$ is the 4-position of the particle at
the reference observable $s$ and,
when moving with 4-momentum $p_\mu$, 
$x^\mu-{p^\mu\over m}s$ is the 4-position at $s=0$.
Therefore, with respect to the reference observable,
(\ref{eq:psi_R_H}) is in the Heisenberg picture 
while (\ref{eq:psi_R_S}) is in the Schr\"odinger picture
before imposing the constraint.
However, after imposing the constraint, (\ref{eq:psi_R_S}) cannot be
in the Schr\"odinger picture since the $s$ dependence in (\ref{eq:psi_R_S})
disappears.
Nevertheless, the Heisenberg picture of (\ref{eq:psi_R_H}) holds 
because the presence of $s$ in (\ref{eq:psi_R_H}) is not due to
the state function itself but due to the gauge dependence of $x^\mu$.
Hence, the reference observable $s$ behaves as if it is a time in 
the Heisenberg picture.
The overall phase factor $e^{-i{m\over 2}s}$ does not affect physics.


\section{General relativity (GR)}\label{sec:gen}
\subsection{Gauge covariant formulation}\label{sub:gr-gauge}

The idea developed in the previous sections for NP and RP is applied to GR.
The three theories are Hamiltonian constraint systems, that is, 
gauge covariant formulations in our classification.
The main difference of GR from NP and RP is that GR is a field theory while
NP and RP are trajectory theories.
The gauge parameter for GR is at least 4 dimensional while it is one dimensional 
for NP or RP.

The action is 
\begin{eqnarray}
&&
S_G:=\int\left(\int P(x){dQ(x)\over dt}d^3x-H_G\right)dt,
\label{eq:S_G}
\end{eqnarray}
where $Q$ and $P$ are the canonical variables, whose physical meaning and
number of indices depend on the formulation of GR.
Here, we suppress possible indices from the canonical variables for simplicity
of notation.
The integrals are over the spatial part and time of spacetime coordinates 
$x^\mu=(t,x^a)$, and 
\begin{eqnarray}
&&
H_G:=\int N^\mu C_\mu d^3x.
\label{eq:H_G}
\end{eqnarray}
is the Hamiltonian.
$N^0$ and $N^a (a=1,2,3)$ are respectively the lapse and shift functions.
$C_0=0$ and $C_a=0$ are known as Hamiltonian and
diffeomorphism constraints respectively.
The coordinates $x^\mu$ do not present physical spacetime but
are arbitrary coordinates. 

The canonical variables, $Q$ and $P$, satisfy the Poisson bracket relations 
\par\noindent
$\{Q(x),P(y)\}=\delta^3(x,y)$ and
$\{Q(x),Q(y)\}=\{P(x),P(y)\}=0$.
They are functions of $x^\mu$ and their evolution in $t$
is generated by the Hamiltonian $H_G$, which is constrained to vanish.
In addition, their diffeomorphisms in $x^a$ are generated by 
the diffeomorphism constraints $C_a$ in the Hamiltonian.

The gauge covariant equations of motion 
and the constraints are formally
\begin{eqnarray}
&&
{df(x)\over dt}=\int d^3y N^\mu(y)\left\{f(x),C_\mu(y)\right\},
\label{eq:dfdt}
\\
&&
C^\mu(x)=0,
\label{eq:cmu}
\end{eqnarray}
where $f(x)$ is $Q(x)$ or $P(x)$.
We refer the subspace in the phase space satisfying the constraints
(\ref{eq:cmu}) to the constraint surface.

\subsection{Spacetime reference observables}\label{sub:spacetime}

In this subsection, we define reference observables.
Then, we derive the equations of motion with respect to them,
that is, what we think gauge independent equations of motion
although they are still formal here.

We define reference observables and denote them by $s^\mu$.
They are observables satisfying  ${ds^\mu\over dt}=N^\mu$ 
on the constraint surface.

The canonical variables are functions of $x^\mu$.
We restrict ourselves to a sector in which the canonical variables 
are functions of $s^\mu$, through which they implicitly depend on $x^\mu$,
that is, $f(x):=f(s(x))$.
If $\vert{\partial s^\mu\over\partial x^\nu}\vert>0$,
then the sector is identical to the space of the canonical variables
we started with.

We use
${d\over dt}={ds^\mu\over dt}{\partial\over \partial s^\mu}
=N^\mu {\partial\over \partial s^\mu}$ 
or more explicitly
\begin{eqnarray}
&&
{df(x)\over dt}={ds^\mu(x)\over dt}{\partial f(x)\over\partial s^\mu}=
\int d^3y N^\mu(y)\delta^3(y,x){\partial f(x)\over\partial s^\mu}
\label{eq:dfxdt}
\end{eqnarray}
to rewrite the gauge covariant equations of motion as follows.
\begin{eqnarray}
&&
{\partial f(x)\over \partial s^\mu}=\int d^3y\{f(x),C_\mu(y)\}.
\label{eq:dfdsmu}
\end{eqnarray}
Here, we have used the fact that $N^\mu$ is arbitrary.

The reference observables for NP and RP were known as physical time
and the proper time respectively.
However, we do not know what the reference observables $s^\mu$ for GR are.
Details are left for future work.

\subsection{Spacetime Heisenberg picture}\label{sub:gr-heise}

Quantization is a process of replacing $P(x)$ by 
$\hat P(x):=-i{\delta\over\delta Q(x)}$
and the Poisson braket by $(-i)$ times the commutator in a representation 
space in which $Q(x)$ is diagonal.

The state function is a functional of $Q$ and $s^\mu$.
The physical state suggested by NP and RP is
\begin{eqnarray}
&
\Psi_G[Q]
&
=e^{-i\int\hat C_\mu(x)s^\mu(x)d^3x}e^{i\int P(x)Q(x)d^3x}
=e^{i\int\left(P(x)Q(x)-C_\mu(x)s^\mu(x)\right)d^3x}
\label{eq:psi_G_S}
\\
&&
=e^{i\int d^3x\int_0^{P(x)}\left(Q(x)
-\int d^3y{\delta C_\mu(y)\over\delta P(x)}s^\mu(y)\right)dP(x)}
\label{eq:psi_G_H}
\end{eqnarray}
with $C_\mu=0$.
Here, the operator applied makes an evolution in $s^\mu$.
$Q(x)$ is the value of $Q$ at the reference observables $s^\mu(x)$ and,
when propagating with $P(x)$, 
$Q(x)-\int d^3y{\delta C_\mu(y)\over\delta P(x)}s^\mu(y)$
is the value of $Q$ at the reference observables $s^\mu(x)=0$.
Therefore, with respect to the reference observables, (\ref{eq:psi_G_H})
is in the Heisenberg picture while (\ref{eq:psi_G_S}) is in the Schr\"odinger
picture before imposing the constraints.
However, after imposing the constraints, (\ref{eq:psi_G_S}) cannot be
in the Schr\"odinger picture since the $s^\mu$ dependence in (\ref{eq:psi_G_S})
disappears.
Nevertheless, the Heisenberg picture of (\ref{eq:psi_G_H}) holds
because the presence of $s^\mu$ in (\ref{eq:psi_G_H}) is not due to the state
function itself but due to the gauge dependence of $Q(x)$.
Perhapes, the reference observables $s^\mu$ might behave as if they are 
spacetime in the Heisenberg picture.


\section{Conclusion}\label{sec:con}

We have studied the Hamiltonian constraint system using simple models 
(NP and RP).
In order to study the time evolution, we defined reference observables,
with respect to which other observables evolve.
The definition was general enough for the Hamiltonian constraint system,
including GR.

{}First, we studied NP, which already had physical time.
We derived equations of motion with respect to the reference observable.
It was known that the theory was well described in terms of the physical time.
Then, we understood that the reference observable we defined for NP was in fact
the physical time.
Upon quantization, we constructed a Heisenberg picture with respect to 
the reference observable.

Next, we studied RP, which did not have Lorentz invariant time, 
without defining one.
We derived equations of motion with respect to the reference observable.
It was known that the theory was well described in terms of the proper time.
Then, we understood that the reference observable we defined for RP was in fact
the proper time.
Upon quantization, we constructed a Lorentz invariant Heisenberg picture 
with respect to the reference observable.

{}Finally, we outlined an application of the idea to GR, which had no time.
Its time coordinate was not physical time.
{}For GR, a field theory, a set of reference observables were defined.
We derived equations of motion with respect to the set of reference observables.
Upon quantization, we constructed a Heisenberg picture with respect to them.
However, we did not know what the reference observables we defined fo GR were.
Details were left for future work.


\end{document}